\documentstyle[11pt,aaspp4]{article}

\def\eg{{\it e.g.}}
\def\et{{\it et al.}}

\def\a{$\&$ }
\def\x{$\times$}
\def\about{$\sim$}

\def\simgt{\buildrel{>}\over \sim}
\def\deg{$^{\rm o}$}
\def\asec{$''$}

\def\mass{${\cal M}$}
\def\msun{${\cal M}_{\odot}$}
\def\sun{$_{\odot}$}
\def\yr-1{yr$^{-1}$}
\def\u{$U_{336}$}
\def\b{$B_{439}$}
\def\v{$V_{555}$}
\def\i{$I_{814}$}
\def\ub{$U_{336}-B_{439}$}
\def\bv{$B_{439}-V_{555}$}
\def\vi{$V_{555}-I_{814}$}

\def\MV{M$_V$}

\def\ha{H$\alpha$}

\def\HST{{\it HST}}

\begin{document}

\title{Cataclysmic Variables and a New Class of Faint UV Stars in the\nl 
Globular Cluster NGC 6397\footnote{Based on observations with
the NASA/ESA {\it Hubble Space Telescope}, obtained at the Space
Telescope Science Institute, which is operated by AURA, Inc., under
NASA contract NAS 5-26555.}}

\author{Adrienne M.\ Cool}

\affil{Department of Physics and Astronomy, San Francisco State University\nl 
1600 Holloway Avenue, San Francisco, CA 94132 \nl
Electronic mail: cool@sfsu.edu}

\author{Jonathan E. Grindlay}

\affil{Department of Astronomy, Harvard University, 60 Garden St.\nl 
Cambridge, MA 01238 \nl
Electronic mail: josh@cfa.harvard.edu}

\author{Haldan N.\ Cohn and Phyllis M. Lugger}

\affil{Department of Astronomy, Indiana University, Swain West\nl 
Bloomington, IN 47405 \nl
Electronic mail: cohn@indiana.edu, lugger@indiana.edu}

\author{Charles D. Bailyn}

\affil{Department of Astronomy, Yale University, P.O. Box 208101\nl 
New Haven, CT 06520-8101\nl
Electronic mail: bailyn@astro.yale.edu}

\begin{abstract}

We present evidence that the globular cluster NGC 6397 contains two
distinct classes of centrally-concentrated UV-bright stars.
Color-magnitude diagrams constructed from \u, \b, \v, and \i\ data
obtained with the \HST/WFPC2 reveal seven UV-bright stars fainter than
the main-sequence turnoff, three of which had previously been
identified as cataclysmic variables (CVs).  Lightcurves of these stars
show the characteristic ``flicker'' of CVs, as well as longer-term
variability.  A fourth star is identified as a CV candidate on the
basis of its variability and UV excess.  Three additional UV-bright
stars show no photometric variability and have broad-band colors
characteristic of B stars.  These non-flickering UV stars are too
faint to be extended horizontal branch stars.  We suggest that they
could be low-mass helium white dwarfs, formed when the evolution of a
red giant is interrupted, due either to Roche-lobe overflow onto a
binary companion, or to envelope ejection following a common-envelope
phase in a tidal-capture binary.  Alternatively, they could be
very-low-mass core-He-burning stars.  Both the CVs and the new class
of faint UV stars are strongly concentrated toward the cluster center,
to the extent that mass segregation from 2-body relaxation alone may
be unable to explain their distribution.

\end{abstract}


\clearpage

\section{Introduction}

The center of NGC 6397 is a prime region in which to investigate the
effects of stellar interactions.  At just 2.2 kpc, its high-density
core can be studied in detail: with \HST, stars well below the turnoff
can be observed all the way to the cluster center.  The cluster's
structure has been investigated most recently by Sosin (1997), who
finds within the power-law cusp a resolved core with a radius of
\about 5\asec.  Fokker-Planck models constructed by Dull (1996) match
the strong mass segregation seen by King, Sosin, \&\ Cool (1995), and
put the central density at \about 6 \x 10$^7$ \msun/pc$^{-3}$.  Dark
remnants in the form of neutron stars and/or massive white dwarfs
appear to make up \about 30\% of the cluster mass (Dull 1996, Drukier
1995).

There is already evidence for activity associated with stellar
interactions in NGC 6397.  Auri\` ere, Ortolani, \&\ Lauzeral (1990)
suggested that the unusually bright blue stragglers within r $<$
5\asec\ of the cluster center could be collisional mergers or
collisionally-hardened primordial binaries.  Hardened primordial
binaries were also put forward by Djorgovski \et\ (1991) as a
potential explanation for the central deficit of red giants.  Three
candidate cataclysmic variables (CVs), the possible products of tidal
capture (\eg, Di Stefano \&\ Rappaport 1993), were found within \about
10\asec\ of the cluster center using \ha\ imaging with \HST/WFPC1
(Cool \et\ 1995), and confirmed spectroscopically with \HST/FOS
(Grindlay \et\ 1995).  Two other faint UV stars near the center were
identified as CV candidates with \HST/FOC (De Marchi \&\ Paresce 1994,
Cool \et\ 1995), but were too faint for the WFPC1 \ha\ study.

Here we present results of an \HST/WFPC2 study of the center of NGC
6397, preliminary results of which have appeared in Cool (1997) and
Sosin (1997).  This paper will focus on the cluster's faint UV stars,
presenting lightcurves, magnitudes, and broad-band colors for the five
previously known objects, together with two newly discovered objects.
A companion paper (Edmonds \et\ 1998; hereafter EG98) presents our
\HST/FOS results for these objects.  Results concerning main-sequence
stars, the white-dwarf sequence, and detached binaries will appear
elsewhere.  We describe the data and results in $\S$2 and discuss the
findings in $\S$3.

\section{Observations and Results}

On 1996 March 6--7 we observed NGC 6397 using the WFPC2, with the PC1
chip approximately centered on the cluster core.  A total of 55
\about 400 s exposures were taken spanning \about 16 hours.  The first
35 exposures used the F336W (``\u'') filter, and the remaining 20 used
the F439W (``\b'') filter.  Shorter exposures were also taken in these
filters and in the F555W (``\v'') and F814W (``\i'') filters.  The
STScI ``pipeline'' calibration procedures were applied to all the
images.  Sets of coaligned \u\ and \b\ images were stacked to produce
deep images free of cosmic rays at each pointing.  Photometry was
carried out on the combined images using the ``weighted, subtracted
aperture photometry'' method described by Cool \&\ King (1995).
Instrumental magnitudes were transformed to the WFPC2 ``synthetic
system'' using the prescriptions in Holtzman \et\ (1995).  Details
will appear in a future paper (Cool \et\ 1998).

The \ub\ vs.~\u\ color-magnitude diagrams (CMDs) are shown in Fig.~1.
The bend in the upper main sequence is a filter effect; the
turnoff is off the top edge of the diagrams.  Stars to the left and
right of the main sequence are largely field stars, owing to the low
galactic latitude of NGC 6397 (b~$=-11$\deg).  The upper end of a
white dwarf (WD) sequence is also visible in the bottom left corner of
each panel.  The stars of interest here are marked with squares and
triangles.  All seven appear in the PC1 chip.  The absence of any
stars in the WF chips in the corresponding region of the CMD implies
that these seven stars are cluster members, given the much larger
combined area of the WF chips relative to the PC1 (a factor of \about
14).  Moreover, the stars are centrally concentrated within the
cluster: each of the four chips contains a similar number of cluster
stars, yet these UV-bright stars are all on the PC1 chip.  Five of the
seven have been identified in previous \HST\ studies as emission-line
stars and/or as UV-bright stars (Grindlay \et\ 1995, and references
therein).  The two new UV-bright stars identified here were either
outside of the field of view or below the magnitude limit of previous
studies.

Lightcurves for the seven UV-bright stars and three non-variable
reference stars are presented in Fig.~2.  Four of the UV-bright stars
are variables.  The three previously-known CVs (CV \#1--3) all show
variability on time-scales of hours, as well as the shorter time-scale
variability known as ``flicker.''  The flicker is most evident for the
faintest of the three (CV \#3), but can be seen even in the brightest
(CV \#1), when compared to a non-variable reference star of similar
magnitude (Ref \#1).  The fourth variable star, whose lightcurve is
very similar to that of CV \#3, we identify as a new candidate CV (CV
\#4 in Fig.~2); follow-up \HST/FOS spectroscopy (EG98) reveals Balmer,
He I, and He II emission lines, confirming it as a CV.  These
flickering variables are marked with triangles in Fig.~1.  The new CV
is the faintest of the four, and was just beyond the detection limit
in our WFPC1 \ha\ study (Cool \et\ 1995).

In contrast to the CVs, the UV-bright stars marked with squares in
Fig.~1 show no sign of photometric variability.  We refer to them
hereafter as ``non-flickerers'' (NFs), and have labeled them NF \#1--3
in Fig.~2.  The rms variations in the lightcurves of these stars are
typical of other non-variable stars of similar apparent magnitude
(cf. Refs \#1--3).

In Fig.~3 we compare the broad-band colors of the CVs vs.\ NFs.  In
the left panel (\ub), all seven stars lie to the left of the main
sequence.  In the central panel (\bv), the brightest two CVs have
shifted onto the main sequence, while the fainter two have moved to
the red side of the three NFs.  The separation between flickerers and
non-flickerers becomes even more distinct in the right-hand panel
(\vi).  Here, all the CVs are either on or very close to the main
sequence.  In contrast, the NFs remain well to the blue side of the
main sequence in every filter combination.  In short, the three NFs
are simply blue stars, whereas the four CVs have UV excesses.  Table 1
lists positions, mean absolute magnitudes and dereddened colors of the
seven faint UV stars, together with their radial offsets from the
cluster center, and a cross reference to the ID numbers used by Cool
\et\ (1995).

\section{Discussion}

The NFs appear to constitute a distinct class of faint UV stars in the
cluster.  Their broad-band colors and lack of photometric variability
both set them apart from the CVs.  In addition, two of the three,
while close to the cluster center, are outside the error circles of
the three central X-ray sources detected with ROSAT (Cool \et\ 1993).
This is in contrast to the four CVs, all of which are within the ROSAT
HRI X-ray error circles.  Thus, NGC 6397 contains two distinct
populations of centrally-concentrated faint UV stars.  Below we
explore what can be learned about the nature and characteristics of
the NFs and CVs from their magnitudes, colors, numbers, and spatial
distribution within the cluster.

The first possibility to consider is that the NFs are simply faint
``extended horizontal branch'' (EHB) stars, which have been observed
in many globular clusters and are thought to be core-helium-burning
stars with unusually thin hydrogen envelopes.  However, EHB stars are
expected to have \MV\ \about 4--5 even when stripped of nearly their
entire hydrogen envelopes (Dorman 1992), more than 2 magnitudes
brighter than the NFs.  Moreover, NGC 6397 shows no sign of a faint
blue HB extension (Lauzeral \et\ 1992).  If the NFs are He-burning
stars, then their masses would need to be significantly lower than
those of the cluster's HB stars.  A scenario in which low-mass
core-He-burning stars may be produced from the merger of two low-mass
He WDs has been explored extensively by Iben (1990).  With
sufficiently small WD progenitor masses, it may be possible to create
merged systems with masses well below the HB mass.  To explain the
NFs, however, it appears that a combination of rather special
circumstances would be required to produce exclusively very-low-mass
merger products.

A more promising possibility is that the NFs are low-mass helium white
dwarfs.  Masses above \about $0.3-0.4$ \msun\ are ruled out, as they
are predicted to be too blue at the absolute magnitudes of the NFs.
(See Cool, Sosin, \&\ King 1997 for a comparison to the log g $=
7.0-9.0$ WD models of Bergeron, Wesemael, \&\ Beauchamp 1995.)  Below
this mass, the expected colors of WDs are consistent with those of the
NFs.  EG98 report \HST/FOS spectroscopy of NF \# 2 showing it has log
g $\simeq$ 6.25 $\pm$ 1.0 and a continuum temperature consistent with
the values log T$_{eff}$ $\simeq$ 4.3 $\pm$ 0.1 and BC$_V$ $\simeq$
2.0 $\pm$ 0.2 derived here from our measured $(B-V)_0$ value.  The
implied luminosities are log L/L\sun $\simeq$ 0.1 to $-$0.6.  In the
helium WD models of Althaus \&\ Benvenuto (1997), these values
correspond to \mass $\simeq$ 0.25\msun and log g $\simeq 6.0-6.7$,
consistent with the spectroscopic results.  Detailed comparisons with
He WDs models, evolutionary lifetimes, and a discussion of the
apparent lack of fainter He WDs in Fig.~1 are given by EG98.

If indeed the NFs are \about 0.25\msun\ He WDs, we need to consider
the constraints on the formation rate of such systems in NGC 6397.  To
form a He WD, the envelope of a red giant branch (RGB) star must
somehow be removed before the first ascent of the branch is complete.
Two mechanisms by which this might be accomplished are: (1) Roche-lobe
overflow onto a binary companion (Webbink 1975); (2) a collision that
produces a common-envelope system, followed by envelope ejection
(Davies, Benz, \&\ Hills 1991).  In either case, the formation rate is
necessarily linked to post--main-sequence evolution.  An order of
magnitude estimate of the fraction of RGB stars that would need to be
stripped can be made by noting that there are six horizontal branch
(HB) stars within the PC1 chip where the three NFs reside.  While the
lifetimes of He WDs are uncertain (see, \eg, Iben \&\ Tutukov 1986),
if we assume that they are similar to those of HB stars (10$^8$ yrs),
then \about $1 \over 3$ of all RGB stars would need to be stripped.
If binaries alone were responsible for the stripping, the required $1
\over 3$ binary fraction would be compatible with binary fractions in
the range 10--40\% that have been suggested (Mateo 1996, and
references therein).  Binaries resident in the core are likely to be
sufficiently hardened as to make Roche-lobe overflow inevitable (Hut
\et\ 1992).  Binaries formed by tidal capture could also contribute to
the pool of He WD progenitors.  If, on the other hand, stellar
interactions were the sole production route for He WDs, the required
collision rate for RGB stars would be \about 3 \x 10$^{-8}$
\yr-1.  This is an order of magnitude higher than the rate at which
common-envelope systems form via collisions between single stars in
NGC 6397, as estimated by Davies (1992).  However, it is
compatible with the rates he computed for binary-single star
encounters.  Thus, binary evolution and/or collisions involving
binaries seem to provide at least plausible mechanisms for forming He
WDs.  Whatever the production mechanism, the existence of He WDs would
provide an important link to the apparent deficit of RGB stars near
the cluster center (Djorgovski \et\ 1991).

Whether He WDs form from pre-existing binaries or via collisionally
formed binaries, they ought to have binary companions at present.  We
can place limits on the mass of any main-sequence companion
by assuming that the \bv\ colors are due exclusively to a He WD, and
asking whether the \vi\ colors are compatible with a
single-temperature model.  We find that they are, and that
companions with \mass $\simgt$ 0.15 \msun\ can be ruled
out.  Fainter main-sequence stars, or dark companions in the form of
cool WDs or neutron stars, are of course allowed.  We note that
massive remnants would be a natural result of either formation
mechanism, as they appear to be the most common species within the
central few arcseconds of the cluster (Dull 1996), where collisions
and/or exchanges into primordial binaries are likely to occur.

Mass estimates for the CV secondaries can also be gleaned from the
present data.  The locations of the CVs on or near the main sequence
in the right-hand panel of Fig.~3 suggest that the secondaries
dominate in \v\ and \i.  By assuming that the disks make a negligible
contribution in \i, we deduce masses of \about\ 0.69, 0.56, 0.48, and
0.44 \msun\ for the secondary stars in CVs \# 1--4, respectively,
using a mass--M$_I$ relation for stars with [Fe/H] $= -1.9$ kindly
provided by E.\ Brocato based on the models by Alexander \et\ (1997).
Evolutionary effects, and thus departures from apparent main sequence
masses, appear to be small for CVs with orbital periods $<$ 7h (Smith
\&\ Dhillon 1998), as is likely the case for these cluster CVs (EG98);
in any case, the proximity of the CVs to the main sequence in \vi\
suggests that they are unevolved.  That all four secondaries are drawn
from the upper half of the mass range of the main sequence is
interesting, considering that CVs with less massive secondaries should
have been observable (see Fig.~1).  Such a preponderance of massive
secondaries would be expected if the CVs form via stellar
interactions.  However, deeper and more systematic CV searches are
needed before the significance of this can be properly assessed.  At
present we simply note that, relative to field CVs, the secondaries
are rather bright (owing to a combination of low metallicity and high
mass), and that the disks are rather faint (\MV\ \about 8-10.5, when
secondaries are subtracted out).  Disks this faint are not unusual
among field CVs, but that all four should be so faint has interesting
implications for their nature (EG98).

Like the bright blue stragglers (BSs), both the CVs and the NFs are
strongly concentrated towards the cluster center.  This effect is
expected due to mass segregation, since these objects are more massive
than typical cluster members.  However, the strength of the central
concentration is hard to understand.  All 13 of the relevant objects
(bright BSs, CVs and NFs) lie within 16\asec\ of the cluster center.
For comparison, in Fokker-Planck models of NGC 6397 (Dull 1996), only
40\% of the neutron stars fall within this radius.  Binary stars and
merger products are unlikely to be more massive than twice the main
sequence turnoff mass, so the typical mass of these exotic objects
should be similar to the 1.4 \msun\ assumed for the neutron stars.
Thus either these unusual objects are much more massive than we expect
(as may be the case for the brightest BS according to Sills
\& Bailyn 1998), or some process other than mass segregation via
2-body relaxation is at work.  The strong central concentration
provides compelling evidence that the formation of CVs, NFs and bright
BSs is linked to stellar interactions, which are strongly favored
within 5\asec\ of the cluster center.

The existence within a single cluster of three distinct classes of
stars, each of whose formation is likely to be tied to stellar
interactions, presents an opportunity to test the outcomes of these
rare but astrophysically important events.  Constraints on the numbers
and radial distributions of various stellar species in globular
clusters are steadily improving thanks in part to star counts provided
by \HST\ and increasingly sophisticated dynamical modeling.  A
reevaluation of stellar interaction rates that incorporates these new
developments would be valuable.  A census of faint UV stars in other
globular clusters would also be of interest, particularly if carried
out in a manner that clearly distinguishes between CVs and NFs.  The
distinction can be made on the basis of \ha\ imaging, multicolor
broad-band photometry, and/or sensitive variability studies.

\acknowledgments

We thank P. Edmonds, B. Dorman, I. King, R. Saffer, A. Sarajedini, and
C. Sosin for discussions, and E.\ Brocato for providing isochrones for
NGC 6397.  This work was supported in part by NASA grant HST-GO-5929.

\newpage
 

\figcaption[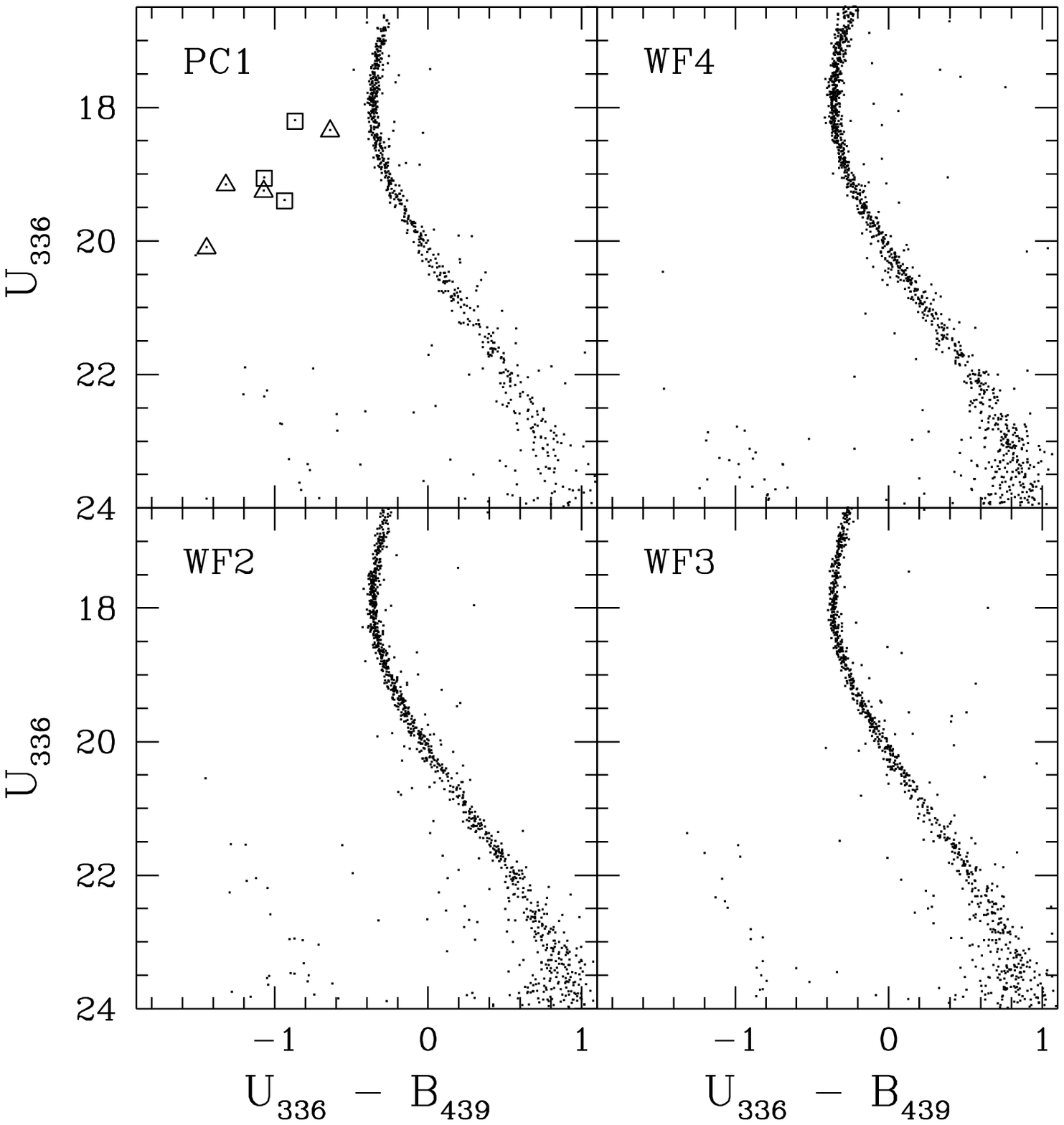]{CMDs for the central regions of NGC 6397,
divided into stars appearing in each of four WFPC2 chips.  Stars
marked with triangles are CVs.  Stars marked with squares are a newly
identified class of non-variable faint UV stars that may be helium
white dwarfs.\label{fig1}}

\figcaption[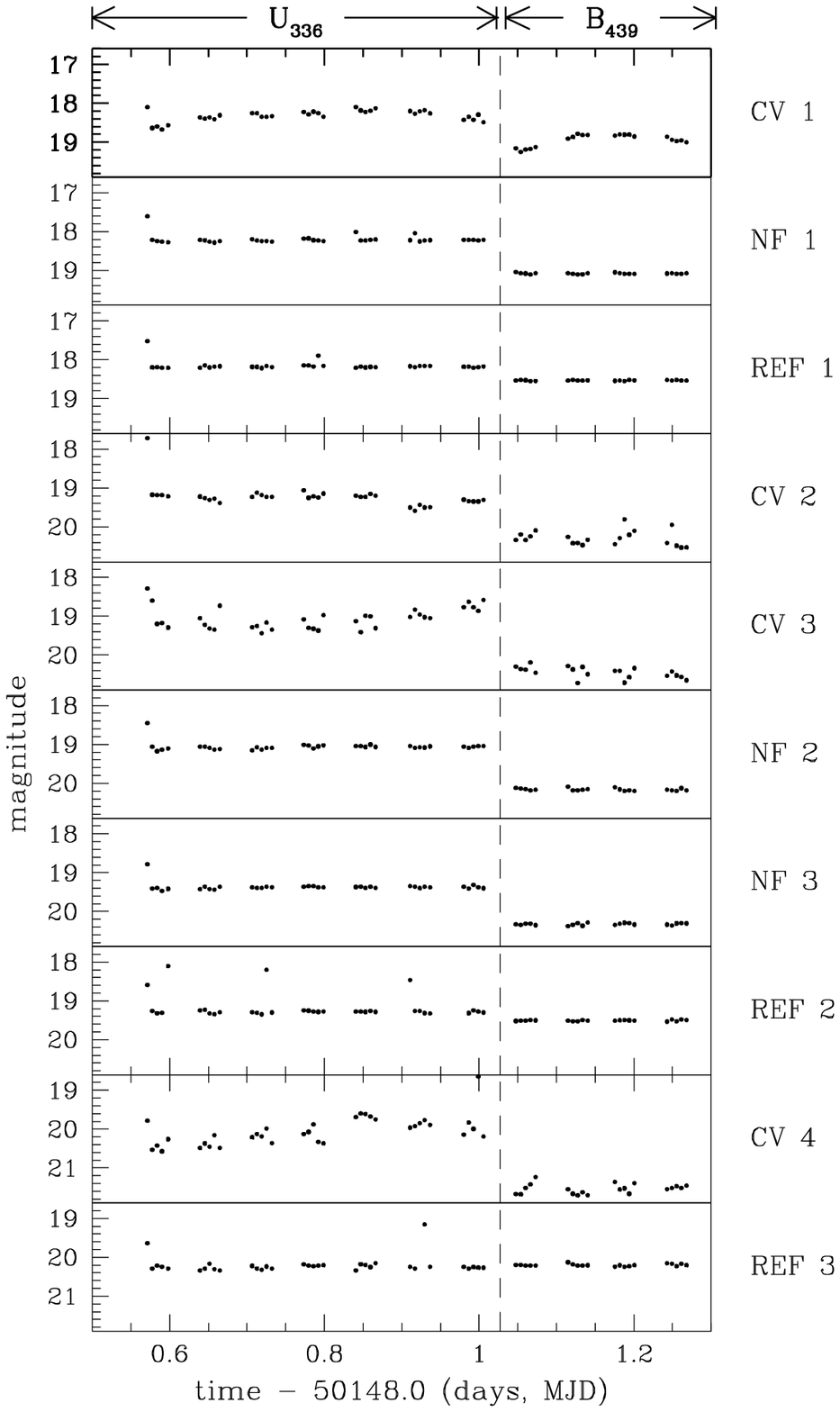]{Lightcurves for the seven faint UV stars near the center
of the cluster, along with three non-variable reference stars.  The
change of filter from \u\ to \b\ part way through the observations is
represented by a dashed vertical line.  Occasional high points (\eg,
$1-4$ per lightcurve), are due to cosmic rays.  The reference stars
provide a baseline for comparison of the UV star lightcurves.  All
four CVs are clearly variable, while the three ``non-flickerers'' have
lightcurves that are very similar to the non-variable reference
stars.\label{fig2}}

\figcaption[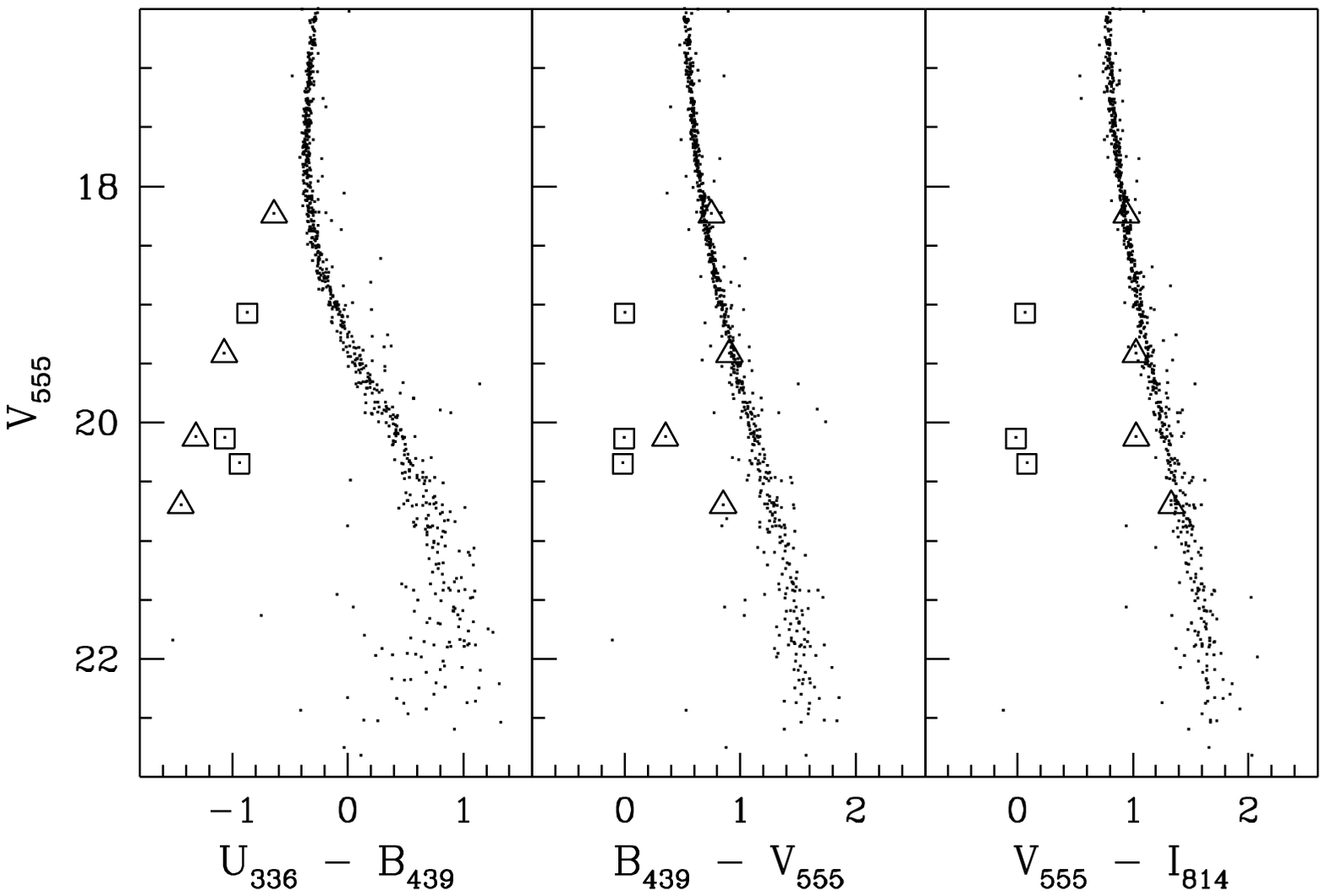]{Three CMDs for stars in the PC1 chip.
Symbols are as in Fig.~1.  Note the differing behavior of the
non-flickerers as compared to the CVs as one moves from the bluest to
the reddest color index.
\label{fig3}}

\newpage

\begin{deluxetable}{cccccccrc}
\tablewidth{40pc}
\tablecaption{Faint UV stars in NGC 6397.}
\tablehead{
\colhead{Star} & \colhead{x} & \colhead{y} & \colhead{\MV} 
& \colhead{$(U-B)_0$} & \colhead{$(B-V)_0$} & \colhead{$(V-I)_0$} 
& \colhead {$\Delta$r (\asec)} & \colhead{ID} 
}
 
\startdata

CV1 & 451 & 589 & 5.95 & $-0.84$ & 0.56    & 0.69    & 10.7 & 1 \nl
CV2 & 513 & 349 & 7.13 & $-1.27$ & 0.72    & 0.77    &  0.9 & 2 \nl
CV3 & 314 & 436 & 7.84 & $-1.51$ & 0.16    & 0.78    &  9.9 & 3 \nl
CV4 & 454 & 525 & 8.41 & $-1.65$ & 0.67    & 1.08    &  7.8 & new \nl
NF1 & 584 & 717 & 6.79 & $-1.06$ & $-0.20$ & $-0.18$ & 16.4 & new \nl
NF2 & 647 & 551 & 7.84 & $-1.27$ & $-0.19$ & $-0.27$ & 10.3 & 7 \nl
NF3 & 442 & 303 & 8.06 & $-1.13$ & $-0.21$ & $-0.17$ &  4.6 & 6 \nl

\enddata

\tablecomments{Star coordinates refer to archive exposure u33r0101t.
For the conversions to absolute magnitudes, we adopted $(m-M)_0 =
11.71$, $A_U = 0.97, A_B = 0.77, A_V = 0.58,$ and $A_I = 0.33$.
Radial offsets are based on a cluster center at (x,y) = (529,378),
following Sosin (1997).}

\end{deluxetable}

\clearpage


\newpage

\begin{references}

\reference{} Alexander, D.\ R., Brocato, E., Cassisi, S., Castellani, V.,
Ciacio, F., \a Degl'Innocenti, S. 1997, \aap, 317, 90
 
\reference{} Althaus, L.~G., \a\ Benvenuto, O.~G. 1997, \apj, 477, 313

\reference{} Auri\` ere, M., Ortolani, S., \a\ Lauzeral, C. 1990,
\nat, 344, 638

\reference{} Bergeron, P., Wesemael, F., \a\ Beauchamp, A. 1995,
\pasp, 107, 1047

\reference{} Cool, A.~M. 1997, in {\it Advances in Stellar Evolution,}
eds.  R.~T.~Rood and A. Renzini (Cambridge: Cambridge U. Press), p.\
191 

\reference{} Cool, A.~M., \et\ 1998, in preparation

\reference{} Cool, A.~M., Grindlay, J.~E., Cohn, H.~N., Lugger, P.~M.,
\a Slavin, S.~D. 1995, \apj, 439, 695

\reference{} Cool, A.~M., Grindlay, J.~E., Krockenberger, M., \a Bailyn,
C.~D. 1993, \apj, 410, L103

\reference{} Cool, A.~M. \a King, I.~R. 1995, in {\it Calibrating
Hubble Space Telescope: Post Servicing Mission,} eds. A.~Koratkar \a
C.~Leitherer (Baltimore: STScI) p. 290

\reference{} Cool, A.~M., Sosin, C., and \a King, I.~R. 1997, in {\it
White Dwarfs,} eds. J. Isern, M. Hernanz and E. Garcia-Berro,
(Dordrecht: Kluwer Academic Publishing), p.\ 129

\reference{} Davies, M.~B. 1992, Ph.D. thesis, Harvard U.

\reference{} Davies, M.~B., Benz, W., \a\ Hills, J.~G. 1991, \apj,
381, 449

\reference{} De Marchi, G. \a\ Paresce, F. 1994, \aap, 281, L13

\reference{} Di Stefano, R. \a Rappaport, S. 1993, \apj, 423, 274

\reference{} Djorgovski, S., Piotto, G., Phinney, E.~S., \a\ Chernoff,
D.~F. 1991, \apj, 372, L41

\reference{} Dorman, B. 1992, \apjs, 81, 221

\reference{} Dull, J.~D. 1996, Ph.D. thesis, Indiana U.

\reference{} Drukier, G.~A. 1995, \apjs, 100, 347

\reference{} Edmonds, P.~D., Grindlay, J.~E., Cool, A.~M., Cohn,
H.~N., Lugger, P.~M., \a\ Bailyn, C.~D., 1998, submitted to \apj\ (EG98)

\reference{} Grindlay, J.~E., Cool, A.~M., Callanan, P.~C., Bailyn,
C.~D., Cohn, H.~N., \a Lugger, P.~M. 1995, \apj, 455, L47 

\reference{} Holtzman, J.~A., Burrows, C.~J., Casertano, S., Hester,
J.~J., Trauger, J.~T., Watson, A.~M., Worthey, G., \pasp, 107, 1065

\reference{} Hut, P. \et\ 1992, \pasp, 104, 981

\reference{} Iben, I. 1990, \apj, 353, 215

\reference{} Iben, I. \a\ Tutukov, A.~V. 1986, \apj, 311, 742

\reference{} King, I.~R., Sosin, C., \a Cool, A.~M. 1995, \apj, 452,
L33

\reference{} Lauzeral, Ortolani, S., Auri\` ere, M., Melnick, J. 1992,
\aap, 262, 63

\reference{} Mateo, M. 1996, in {\it the Origins, Evolution, and
Destinies of Binary Stars in Clusters}, eds. E.~F. Milone and
J.-C. Mermilliod, ASP Conf.\ Series Vol.\ 90, p. 21

\reference{} Sills, A. \&\ Bailyn, C.~D. 1998, in preparation

\reference{} Smith, D.~A. \a Dhillon, V.~S. 1998, \mnras, in press

\reference{} Sosin, C.\ A., 1997, Ph.\ D.\ thesis, U.\ C. Berkeley
 
\reference{} Webbink, R.~F. 1975, \mnras, 171, 555

\end{references}
\end{document}